\documentclass[12pt,thmsa, titlepage]{article}
\usepackage{amssymb}

\usepackage{sw20aip}

\input{tcilatex}
\input epsf
\begin{document}

\author{Ronald Bryan \and Department of Physics, Texas A\&M\ University, \and %
College Station, Texas 77843, U. S. A. \and Telephone: (409) 845-5636;
e-mail: bryan@phys.tamu.edu; \and FAX: 409-845-2590}
\title{Are the Dirac particles of the Standard Model dynamically confined states in
a higher-dimensional flat space?$\thanks{%
PACS: 12.10 Dm Keywords: noncompact eight-dimensional space; soliton;
harmonic-oscillator potential}$}
\date{April 1, 1999}
\maketitle

\begin{abstract}
Some time ago Rubakov and Shaposhnikov suggested that elementary particles
might be excitations trapped on a soliton in a flat higher dimensional
space. They gave as an example $\phi ^{4}$ theory in five dimensions with a
bosonic excitation on a domain wall $\phi _{cl}$ in the fifth dimension.
They also trapped a chiral Dirac particle on the domain wall with the
interaction lagrangian $g\overline{\mathbf{\psi }}\mathbf{\phi \psi }$. The
field equation for the Dirac field was $-i\sum_{\mathbf{\mu }=0}^{4}\mathbf{%
\gamma }^{\mathbf{\mu }}\partial _{\mathbf{\mu }}\mathbf{\psi }-g\mathbf{%
\phi \psi }=0.$ We show that a Dirac equation in \textit{eight} flat
dimensions with $-g\mathbf{\phi }$ replaced by a harmonic-oscillator
``potential'' in the four higher dimensions (plus a large constant $M_{0})$
generates bound states in approximate $SU\left( 4\right) \times SU\left(
2\right) $ representations. These representations have corresponding
``orbital'' times ``spin'' quantum numbers and bear some resemblance to the
quarks and leptons of the Standard Model. Soliton-theory suggests that the
harmonic-oscillator potential should rise only a finite amount. This will
then limit the number of generations in a natural way. It will also mean
that particles with sufficient energy might escape the well altogether and
propagate freely in the higher dimensions. It may be worthwhile to search
for a soliton (instanton?) which can confine Dirac particles in the manner
of the harmonic-oscillator potential.
\end{abstract}

\section{Introduction}

Domain walls and other solitons play a central role in many theories of
elementary particles. In 1983 Rubakov and Shaposhnikov (RS) suggested that
solitons might even constitute an alternative to compactification in
Kaluza-Klein and string models [1]. They gave as an example $\mathbf{\ \phi }
^{4}$-theory in five flat (non-compact) dimensions, choosing 
\begin{equation}
\mathcal{L}=\tfrac{1}{2}\sum_{\mu =0}^{4}\left( \frac{\partial \mathbf{\phi }%
}{\partial \mathbf{x}^{\mathbf{\mu }}}\right) \left( \frac{\partial \mathbf{%
\phi }}{\partial \mathbf{x}_{\mathbf{\mu }}}\right) +\tfrac{1}{2}m^{2}%
\mathbf{\phi }^{2}-\tfrac{1}{4}\lambda \mathbf{\phi }^{4},
\end{equation}
with $\mathbf{\phi }$ a real, one-component field and $\mathbf{g}^{\mathbf{%
\mu \nu }}=$ diag$.\left( +,-,-,-;-\right) .$ If $m^{2}>0,$ then domain-wall
solutions can occur. RS noted that if a domain-wall appears and is assigned
to the fifth dimension, then an excitation of the wall can be interpreted as
a boson free to propagate in the four dimensions of ordinary space-time but
trapped on the wall in the extra dimension. What sets this manner of
confinement apart from ordinary compactification scenarios, of course, is
that a particle of sufficient energy might escape confinement and propagate
freely in all five dimensions.

RS also generated a massless (chiral) Dirac particle by adding to lagrangian
1 the terms 
\begin{equation}
\mathcal{L}_{D}=i\overline{\mathbf{\psi }}\sum_{\mu =0}^{4}\mathbf{\gamma }^{%
\mathbf{\mu }}\partial _{\mathbf{\mu }}\mathbf{\psi }+g\overline{\mathbf{%
\psi }}\mathbf{\phi \psi ,}
\end{equation}
where $\ \mathbf{\gamma }^{\mathbf{\mu }}\mathbf{\gamma }^{\mathbf{\nu }}%
\mathbf{+\;\mathbf{\gamma }^{\mathbf{\nu }}\mathbf{\gamma }^{\mathbf{\mu }%
}=2g}^{\mathbf{\mu \nu }},\;\mathbf{\mu ,\nu }=0,1,\,.\,.\,4.$ \ As in the
case of the boson, the Dirac particle is free to propagate in $M^{4}$ but
skates on the wall in the fifth flat dimension.

Trapping a Dirac particle on an extra flat dimension also plays a role in
lattice gauge theories these days. To avoid Dirac-particle doubling, Kaplan
introduced a fifth flat dimension and generated a massless Dirac particle
with a lagrangian similar to that of Rubakov and Shaposhnikov [2]. When
Kaplan's lagrangian is expressed on a lattice, the second Weyl fermion
appears on an opposite wall in the fifth dimension with exponentially
vanishing overlap with the first Weyl fermion, thus getting around the no-go
theorems [3]. Jansen provides a review of work following Kaplan's original
paper [4].

A question that naturally comes to mind is whether there exists a soliton in
a larger space which can generate \textit{all three generations} of quarks
and leptons. RS remarked that if a domain wall could dynamically confine a
particle in one flat extra dimension, then perhaps a vortex could confine it
in two, a monopole in three, and an instanton in four flat extra dimensions.
Could the mass-spectrum of such a confined particle exhibit the quantum
numbers of the quarks and leptons?

For our part, we had carried out a study of the Klein-Gordon equation
extended to eight flat dimensions $M^{4}\times \widetilde{R}^{4}$ and found
that if we included a symmetrical harmonic-oscillator term in the four extra
dimensions, then we obtained confined $SU\left( 4\right) $ solutions
suggestive of the three generations of quarks and leptons [5]. However a
physical basis for the harmonic-oscillator term was lacking. Could a soliton
(instanton?) provide the confinement?

In the remainder of this paper, we propose a Dirac field equation in eight
flat dimensions which generates Dirac particles with quantum numbers
suggestive of quarks and leptons. We will indicate how these particles might
be coupled to gluons and the electroweak bosons\footnote{%
Dvali and Shifman have proposed a mechanism for localizing massless \emph{\
gauge} bosons on a domain wall [6]; they illustrate this with a domain wall
on the $(x,y)$ plane in $M^{4}$.} in a manner consistent with a left-right
symmetric extension of the Standard Model [7].

\section{Domain walls in five dimensions}

\smallskip To motivate the extension to eight dimensions, let us review the $%
\mathbf{\phi }^{4}$-model in five dimensions. Lagrangian 1 generates the
field equation\footnote{%
Symbols relating to $M^{4}$($\tilde{R}$ , $\,M^{4}\times \tilde{R}$ ) will
be printed normally (with a tilde, in \textbf{bold-face type}). $\tilde{R}$
refers to either $\tilde{R}^{1}$ or $\tilde{R}^{4}$.} 
\begin{equation}
\sum_{\mathbf{\mu }=0}^{4}\partial ^{\mathbf{\mu }}\partial _{\mathbf{\mu }}%
\mathbf{\phi }-m^{2}\mathbf{\phi }+\lambda \mathbf{\phi }^{3}=0
\end{equation}
which admits the ``kink'' solution 
\begin{equation}
\mathbf{\phi =}\left( m/\sqrt{\lambda }\right) \tanh \left[ m\left( 
\widetilde{x}-\widetilde{x}_{0}\right) /\sqrt{2}\right] \equiv \widetilde{%
\phi }_{cl}.
\end{equation}
(Here we denote the fifth dimension $x^{4}\equiv \widetilde{x},$ and set $%
\widetilde{x}_{0}$ equal to zero.) $\widetilde{\phi }_{cl}$ minimizes the
action locally, and if $\mathbf{\phi }$ is expanded about it, then$\;\mathbf{%
\eta \equiv \phi -}\widetilde{\phi }_{cl}$ satisfies 
\begin{equation}
\left( \sum_{\mathbf{\mu }=0}^{4}\partial ^{\mathbf{\mu }}\partial _{\mathbf{%
\mu }}-m^{2}+3\lambda \widetilde{\phi }_{cl}^{2}\right) \mathbf{\eta }=0;
\end{equation}
terms cubic and quartic in $\mathbf{\eta }$ are to be treated by standard
perturbation theory. If we set $\mathbf{\eta =}\eta \left( x^{\mu }\right) 
\widetilde{\eta }\left( \widetilde{x}\right) ,$ then 
\begin{equation}
\left( \square +M^{2}\right) \eta =0
\end{equation}
where $\square $ is the d'Alembertian in $M^{4}$, and 
\begin{equation}
\left( -\partial ^{2}/\partial \widetilde{x}^{2}+3\lambda \widetilde{\phi }%
_{cl}^{2}-m^{2}\right) \widetilde{\eta }=M^{2}\widetilde{\eta }.
\end{equation}
$M$ is taken to be the scalar boson's mass. Eq. 7 is a one-dimensional
Schr\"{o}dinger-like equation that admits a massless solution (which merely
corresponds to a translation of the soliton), a solution $\widetilde{\eta }%
_{3/2}=\sinh \widetilde{z}/\cosh ^{2}\widetilde{z}$, ($\tilde{z}\equiv m%
\tilde{x}/\sqrt{2}$) with $M^{2}=\frac{3}{2}m^{2}$ which represents a boson
trapped in the well, and continuum solutions $\widetilde{\eta }_{c}$ for $%
M^{2}\geq 2m^{2}$ which represent bosons free to propagate in all five
dimensions [1, 8]. We have plotted $\widetilde{\eta }_{3/2}$ in Fig. 1 along
with the confining potential $3\lambda \widetilde{\phi }_{cl}^{2}-m^{2}%
\equiv \tilde{V}$.

\begin{figure}[h]
\epsfxsize=13cm \centerline{\epsfbox{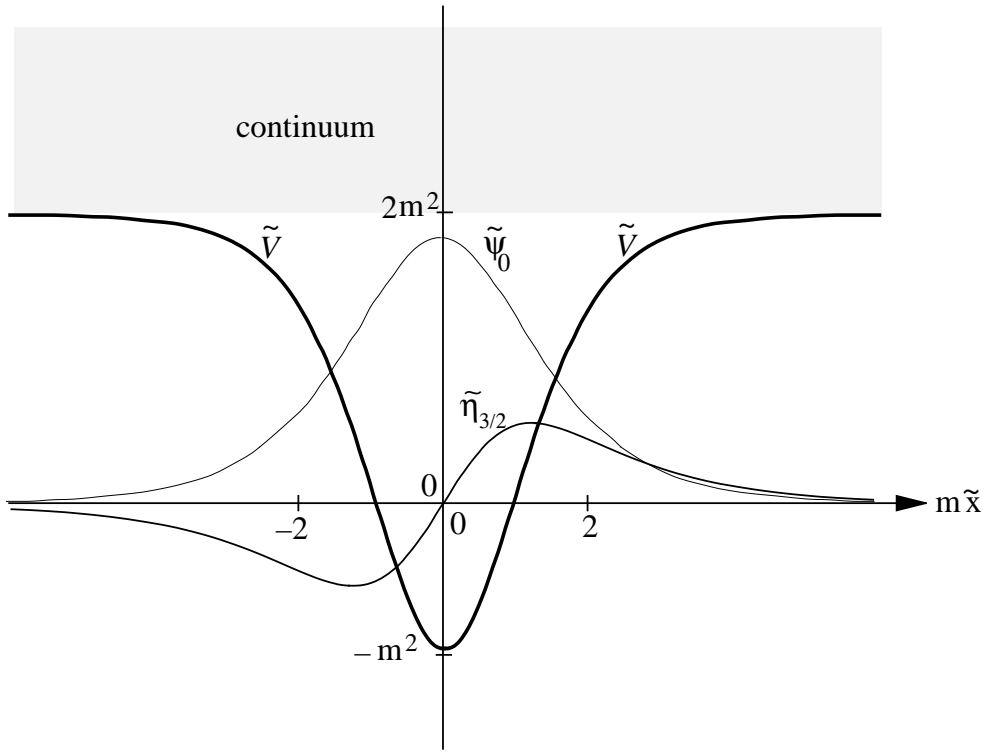}}
\caption{Confining potential $\widetilde{V}$ as a function of the extra
dimension $\widetilde{x}$;$\;$\newline
$\widetilde{V}$ is generated by domain-wall solution $\widetilde{\phi }_{cl}$%
. Also shown is the wavefunction $\widetilde{\eta }_{3/2}$ of the boson
trapped in the well, and the wavefunction $\widetilde{\psi }_{0}$ of a
massless Dirac particle trapped on the wall, for $g=\protect\sqrt{\lambda }.$%
}
\end{figure}

If lagrangian 1 is augmented by lagrangian 2, then minimizing the action
yields the field equation 
\begin{equation}
-i\sum_{\mathbf{\mu }=0}^{4}\mathbf{\gamma }^{\mathbf{\mu }}\partial _{%
\mathbf{\mu }}\mathbf{\psi }-g\mathbf{\phi \psi }=0.
\end{equation}
If $\mathbf{\phi }$ is approximated by the domain-wall solution $\widetilde{
\phi }_{cl}$, then 
\begin{equation}
-i\sum_{\mathbf{\mu }=0}^{4}\mathbf{\gamma }^{\mathbf{\mu }}\partial _{%
\mathbf{\mu }}\mathbf{\psi }-g\widetilde{\phi }_{cl}\mathbf{\psi }=0.
\end{equation}
This equation admits a solution $\mathbf{\psi =}\psi _{0,L}\widetilde{\psi }
_{0}$, where $\psi _{0,L}$ is a left-helical, massless wavefunction in $%
M^{4} $ and 
\begin{equation}
\widetilde{\psi }_{0}=\left( \cosh m\widetilde{x}/\sqrt{2}\right) ^{-g\sqrt{
2/\lambda }}.
\end{equation}
$\widetilde{\psi }_{0}$ is plotted in Fig 1 for the case $g=\sqrt{\lambda }.$
(There is no massless right-helical solution.) There are also unconfined
Dirac states of mass $\geq $ $gm/\sqrt{\lambda }.$

\section{Eight-dimensional Dirac equation}

Let us suppose that there exists a generalization of Eq. 9 in a
higher-dimensional flat space of sufficient dimensionality to accommodate
all of the quarks and leptons in the Standard Model. If so, then how many
extra dimensions are required, and what kind of ``potential'' is needed? The
difficulty is that $\widetilde{N}$-dimensional extensions naturally lead to $%
SO(\widetilde{N})$-symmetry groups, whereas quarks require $SU(\widetilde{N}
) $-type symmetry. However, if the potential generates \textit{%
harmonic-oscillator} states, then these states \emph{will} be
representations of an $SU(\widetilde{N})$-algebra. (See Appendix A.). Since
quarks exhibit $SU(3)$ color-symmetry, this would be a vote for $\widetilde{N%
}=3$ and a symmetrical harmonic-oscillator potential. Because the
coordinates are real, only the (triangular) \textbf{1}, \textbf{3}, \textbf{6%
}, \textbf{10}, . . . \textbf{\ \ }representations would appear, each just
once [5]. The \textbf{3} might correspond to a quark. (The \textbf{1} might
correspond to a ``lepton''. The \textbf{6, 10, . .} would be new species of
Dirac particles.) The spin-degree of freedom might simulate weak isospin.

To accommodate all three generations of quarks and leptons, a higher
symmetry would be required. $SU\left( 4\right) $ might suffice. For four
(real) extra coordinates, the $SU\left( 4\right) $-wavefunctions have the
tetrahedral \textbf{1}, \textbf{4}, \textbf{10}, \textbf{20}, . .
representations, with the \textbf{4} breaking to \textbf{1}$+$\textbf{3},
the \textbf{10} breaking to \textbf{1}$+$\textbf{3}$+$\textbf{6}, \textit{etc%
}. The $SU\left( 4\right) $-\textbf{1} might represent a first-generation
``lepton'', the \textbf{1}$+$\textbf{3 = 4} a second-generation ``lepton''
and a first-generation ``quark'', \textit{etc}. Weight diagrams of the
lowest $SU\left( 4\right) $-multiplets are sketched in Fig. 2.

\begin{figure}[p]
\epsfysize=12cm \centerline{\epsfbox{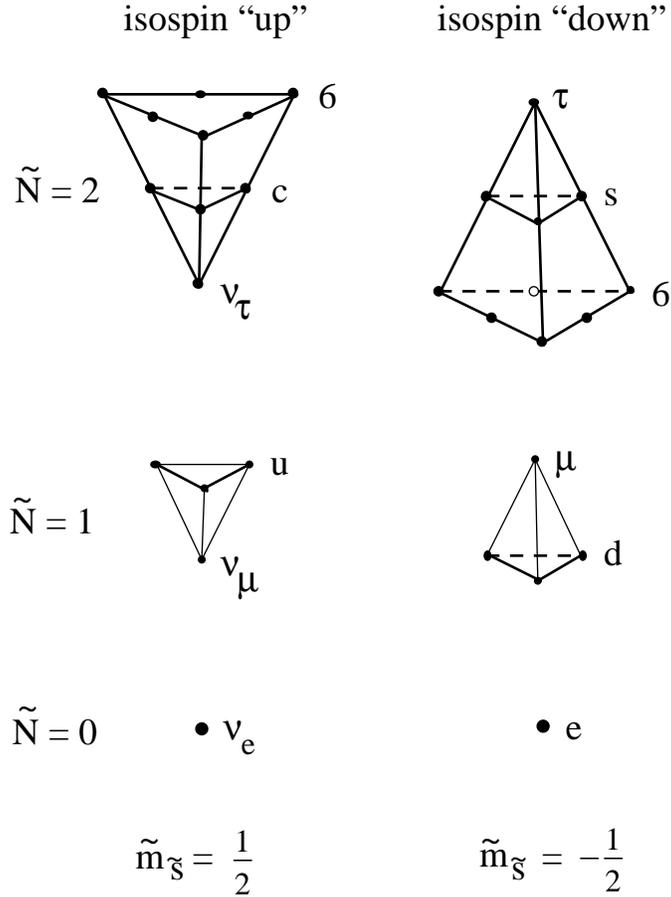}}
\caption{Weight diagrams of lowest lying $SU\left( 4\right) \times SU\left(
2\right) $ multiplets of Eqs. 20 and 24. Diagrams plotted \textit{vs}.
principal $SU\left( 4\right) $ quantum number $\widetilde{N}$ and $SU\left(
2\right) $ quantum number $\tilde{m}_{\tilde{s}}$ . ``Quarks'' and
``leptons'' suggested by these quantum numbers are indicated. All members of
a given multiplet are degenerate in mass. However the weight diagrams are
oriented so that heavier particles appear above lighter particles; \emph{i.e.%
}, the $\widetilde{n}_{4}$ (``generation-number'') axis points downward for $%
\tilde{m}_{\tilde{s}}=\frac{1}{2}$ multiplets and upward for $\tilde{m}_{%
\tilde{s}}=-\frac{1}{2}$ multiplets.}
\end{figure}

Could a harmonic-oscillator ``potential'' actually exist in a higher
dimensional flat space? Possibly. Dvali and Shifman have shown that for some
topological defects, in supersymmetric theories at least, such
oscillator-type potentials appear naturally [9].

However a Dirac equation with a harmonic-oscillator potential will not
necessarily generate harmonic-oscillator wavefunctions. A Schr\"{o}dinger
equation or a Klein-Gordon equation will [5], but a Dirac equation might not
because the Dirac operator ``squares'' the potential\footnote{%
Mathematically speaking, a Dirac equation in four Euclidian dimensions
cannot be $SU\left( 4\right) $-symmetric because it incorporates just four $%
\tilde{\gamma}^{\tilde{\mu}}$-matrices, and the corresponding generators $%
\tilde{\gamma}^{\tilde{\mu}}\tilde{\gamma}^{\tilde{\nu}},$ $\mu ,$ $\nu =1,$ 
$2,\;3,\;4$ can only constitute an $SO\left( 4\right) $-algebra. Adding a
potential will not alter this fact. To generate an $SU\left( 4\right) $
-algebra requires \textit{eight} Dirac matrices [10] such as appear in the
set of $SU\left( 4\right) $-generators $\tilde{\theta}^{\tilde{\mu}^{\dagger
}}\tilde{\theta}^{\tilde{\nu}},$ $\;\mu ,$ $\nu =1,\;2,\;3,\;4,$ where $%
\tilde{\theta}^{\tilde{\mu}^{\dagger }}=\frac{1}{2}\left( -i\tilde{\gamma}^{%
\tilde{\mu}}+\tilde{\gamma}^{\tilde{\mu}+4}\right) $ and $\tilde{\theta}^{%
\tilde{\nu}}=\frac{1}{2}\left( i\tilde{\gamma}^{\tilde{\nu}}+\tilde{\gamma}^{%
\tilde{\nu}+4}\right) .$}. However consider the hydrogen atom in ordinary
space-time$.$ We know that the Coulomb potential in the Dirac equation of
the H-atom reappears as the Coulomb potential in the (Pauli-reduced)
Schr\"{o}dinger equation. This happens because the rest-mass of the electron
is very large compared to the kinetic and potential energies of the electron
and thereby minimizes the symmetry-breaking ``small'' components. Similarly
if we extend Eq. 8 to eight flat dimensions, replace the domain wall with a
symmetrical harmonic-oscillator potential $\mathbf{V}$ in the four extra
dimensions and insert a large ``rest-mass'' $M_{0}$, yielding 
\begin{equation}
\left( -i\sum_{\mathbf{\mu }=0}^{7}\mathbf{\gamma }^{\mathbf{\mu }}\partial
_{\mathbf{\mu }}\;\mathbf{+\;V}+M_{0}\right) \mathbf{\psi }=0,
\end{equation}
then $\mathbf{V}$ reappears unsquared in the two-component reduction of the
Dirac equation in the four higher dimensions. This Pauli equation will
generate exact $SU\left( 4\right) $ harmonic-oscillator ``large''
components, and the symmetry-breaking lower components will be small. We
will show this directly.

Let $\mathbf{\gamma }^{\mathbf{\mu }}\mathbf{\gamma }^{\mathbf{\nu }}+%
\mathbf{\gamma }^{\mathbf{\nu }}\mathbf{\gamma }^{\mathbf{\mu }}=2\mathbf{g}
^{\mathbf{\mu \nu }},$\ $\mathbf{\mu ,\nu }=0,$\ $1,$\ $\;7,$\ where the
metric-tensor $\mathbf{g}^{\mathbf{\mu \nu }}$\ =\ diag.$\left( +,-,-,\ldots
,-\right) .$\ A convenient representation of the gamma-matrices is the
``chiral'' set 
\begin{equation}
\mathbf{\gamma }^{\mathbf{\mu }}=\left\{ 
\begin{array}{ll}
\gamma ^{\mu }\otimes \widetilde{\gamma }^{5}, & \mathbf{\mu }=\mu =0,1,2,3,
\\ 
1\otimes \widetilde{\gamma }^{5}\widetilde{\gamma }^{\widetilde{\mu }}, & 
\mathbf{\mu }=\widetilde{\mu }+3=4,5,6,7
\end{array}
\right.
\end{equation}
where $\gamma ^{\mu },$ $\;\mu =0,\;1,\;2,\;3$ is any standard set of Dirac
matrices while

$\widetilde{\gamma }^{\widetilde{j}}=\left( 
\begin{array}{ll}
\;\;0 & i\widetilde{\sigma }^{\widetilde{j}} \\ 
-i\widetilde{\sigma }^{\widetilde{j}} & 0
\end{array}
\right) ,\;\widetilde{j}=1,2,3,$ $\;\;\;\widetilde{\gamma }^{4}=\left( 
\begin{array}{ll}
0 & 1 \\ 
1 & 0
\end{array}
\right) ,$ \ \ and $\qquad \widetilde{\gamma }^{5}=\left( 
\begin{array}{ll}
1 & \;\,0 \\ 
0 & -1
\end{array}
\right) .$ The $\widetilde{\sigma }^{\widetilde{j}}$ are the $2\times 2$
Pauli spin matrices. The potential $\mathbf{V}=1\otimes \tilde{\gamma}^{5}%
\tilde{V}\left( \tilde{x}\right) $, where $\tilde{V}\ $has the
harmonic-oscillator form.

If one multiplies Eq. 11 by $1\otimes \widetilde{\gamma }^{5}$ and sets $%
\mathbf{\psi }=\psi \left( x\right) \otimes \widetilde{\psi }\left( \tilde{x}%
\right) $ , then (11) separates into 
\begin{equation}
\left( -i\mathbf{\rlap/\nabla }+M\right) \psi \left( x\right) =0
\end{equation}
and 
\begin{equation}
\left( -i\tilde{\rlap/\nabla }+\widetilde{V}+\widetilde{\gamma }%
^{5}M_{0}\right) \widetilde{\psi }=M\widetilde{\psi ,}
\end{equation}
where $\mathbf{\rlap/\nabla }=\sum_{\mu =0}^{3}\gamma ^{\mu }\partial
/\partial x^{\mu }$ and $\tilde{\rlap/\nabla }=\sum_{\widetilde{\mu }=1}^{4}%
\widetilde{\gamma }^{\widetilde{\mu }}\partial /\partial \widetilde{x}^{%
\widetilde{\mu }}$. We have denoted $\left( \mathbf{x}^{4},\mathbf{x}^{5},%
\mathbf{x}^{6},\mathbf{x}^{7}\right) \equiv \left( \widetilde{x}^{1},%
\widetilde{x}^{2},\widetilde{x}^{3},\widetilde{x}^{4}\right) .$ $M$ (not $%
M_{0})$ is taken to be the rest-mass of the Dirac particle in ordinary
space-time$.$ In practice, one solves Eq. 14 for the eigenfunction $%
\widetilde{\psi }$ and eigenmass $M$, then inserts $M$ in Dirac Eq. 13.

The Dirac equation in the extra dimensions may be expanded to 
\begin{equation}
\left( 
\begin{array}{ll}
\widetilde{V}+M_{0}-M & \qquad \widetilde{D} \\ 
\qquad \widetilde{D}^{\dagger } & \widetilde{V}-M_{0}-M
\end{array}
\right) \left( 
\begin{array}{l}
\widetilde{\phi } \\ 
\widetilde{\chi }
\end{array}
\right) \;=\;0,
\end{equation}
where 
\begin{equation}
\widetilde{D}\;=\;\left( 
\begin{array}{ll}
\widetilde{\partial }_{3}-i\widetilde{\partial }_{4}, & \quad \widetilde{
\partial }_{1}-i\widetilde{\partial }_{2} \\ 
\widetilde{\partial }_{1}+i\widetilde{\partial }_{2}, & -\;\widetilde{
\partial }_{3}-i\widetilde{\partial }_{4}
\end{array}
\right)
\end{equation}
with $\widetilde{\partial }_{\widetilde{\mu }}\equiv \partial /\partial 
\widetilde{x}^{\widetilde{\mu }},\;\widetilde{\mu }=1,\;2,\;3,\;4.$ \quad If 
$M_{0}\gg \left\langle \widetilde{V}\right\rangle $ and $M\sim M_{0},$ then 
\begin{equation}
\widetilde{\chi }\thickapprox \frac{1}{2M_{0}}\widetilde{D}^{\dagger }%
\widetilde{\phi },
\end{equation}
yielding

\begin{equation}
\left( \widetilde{V}+M_{0}-M\right) \widetilde{\phi }\,+\,\widetilde{D}
\left( 2M_{0}\right) ^{-1}\widetilde{D}^{\dagger }\widetilde{\phi }=0.
\end{equation}
$\widetilde{D}\widetilde{D}^{\dagger }=\widetilde{D}^{\dagger }\widetilde{D}
=-\;\tilde{\square}\cdot \widetilde{1}_{2\times 2}$ where $\tilde{\square}$
is the four-dimensional laplacian in $\widetilde{R}^{4},$ so Eq.18 reduces
to 
\begin{equation}
\left( -\;\frac{1}{2M_{0}}\tilde{\square}+\widetilde{V}\right) \widetilde{
\phi }=\left( M-M_{0}\right) \widetilde{\phi }.
\end{equation}
Eq.\thinspace 19 is analogous to the Pauli reduction of the time-independent
Dirac equation in ordinary space, with $M-M_{0}$ playing the role of the
kinetic energy. If $\widetilde{V}$ is a symmetrical harmonic-oscillator
potential, then solutions $\tilde{\phi}$ comprise an exact representation of
an $SU\left( 4\right) $-algebra. In particular, if $\widetilde{V}$=$%
\widetilde{\alpha }^{3}\widetilde{X}^{2},$ where $\widetilde{\alpha }$ is an
adjustable real constant and $\widetilde{X}^{2}=\sum_{\widetilde{\mu }
=1}^{4}\left( \widetilde{x}^{\widetilde{\mu }}\right) ^{2},$ then 
\begin{equation}
\tilde{\phi}\,=\,\prod_{\tilde{\mu}=1}^{4}\tilde{\phi}_{\tilde{n}_{\tilde{\mu%
}}}\left( \tilde{\beta}\tilde{x}^{\tilde{\mu}}\right) \,\equiv \,\tilde{\phi}
_{\tilde{n}_{1}\tilde{n}_{2}\tilde{n}_{3}\tilde{n}_{4}}\,\equiv \,\tilde{\phi%
}_{\tilde{n}},
\end{equation}
a product of (unnormalized) harmonic-oscillator functions 
\begin{equation}
\tilde{\phi}_{\tilde{n}_{\tilde{\mu}}}\left( \tilde{\beta}\tilde{x}^{\tilde{%
\mu}}\right) =H_{\tilde{n}_{\tilde{\mu}}}\left( \tilde{\beta}\tilde{x}^{%
\tilde{\mu}}\right) \exp \left[ -\tfrac{1}{2}\tilde{\beta}^{2}(\tilde{x}^{%
\tilde{\mu}})^{2}\right] ,\;\;\;\;\tilde{\mu}=1,\;\,2,\;3,\;4;
\end{equation}
here $\tilde{\beta}^{4}=2M_{0}\tilde{\alpha}^{3}$ and $H_{\tilde{n}_{\tilde{%
\mu}}}$ is a Hermite polynomial of degree $\tilde{n}_{\tilde{\mu}
}=0,\;1,\;2,\;3,\;.\;.$ The associated eigenmass $M=M_{0}+\left( 2\tilde{N}%
+4\right) (\tilde{\beta}^{2}/2M_{0})\equiv \overline{M}^{\left( \tilde{N}
\right) },$ where 
\begin{equation}
\tilde{N}=\tilde{n}_{1}+\tilde{n}_{2}+\tilde{n}_{3}+\tilde{n}
_{4}=0,\;1,\;2,\;3,\;.\;.
\end{equation}
(See Appendix A.)

Actually the solutions to Pauli equation 19 exhibit a higher symmetry than $%
SU\left( 4\right) $ since they also include constant two-component spinors $%
\tilde{\xi}_{\tilde{m}_{\tilde{s}}}={\binom{1}{0}}$ and ${\binom{0}{1}}$
which are representations of $SU\left( 2\right) $.$\ $Thus 
\begin{equation}
\tilde{\phi}\,=\,\tilde{\phi}_{\tilde{n}_{1}\tilde{n}_{2}\tilde{n}_{3}\tilde{%
n}_{4}}\tilde{\xi}_{\tilde{m}_{\tilde{s}}}\,\equiv \,\tilde{\phi}_{\tilde{n},%
\tilde{m}_{\tilde{s}}},
\end{equation}
where the $\tilde{\phi}_{\tilde{n},\tilde{m}_{\tilde{s}}}$ are
representations of $SU\left( 4\right) \times SU\left( 2\right) .$

The solutions to the Dirac equation in the extra dimensions, Eq.\thinspace
14, are thus 
\begin{equation}
\tilde{\psi}=\left( 
\begin{array}{l}
\tilde{\phi} \\ 
\tilde{\chi}
\end{array}
\right) \approx \left( 
\begin{array}{l}
\;\;\;\;\;\;\tilde{\phi}_{\tilde{n},\tilde{m}_{\tilde{s}}} \\ 
\frac{1}{2M_{0}}\tilde{D}^{\dagger }\tilde{\phi}_{\tilde{n},\tilde{m}_{%
\tilde{s}}}
\end{array}
\right) \equiv \tilde{\psi}_{\tilde{n}_{1}\tilde{n}_{2}\tilde{n}_{3}\tilde{n}
_{4}\tilde{m}_{\tilde{s}}}\equiv \tilde{\psi}_{\tilde{n},\tilde{m}_{\tilde{s}%
}},
\end{equation}
with the same eigenmasses $\overline{M}^{\left( \tilde{N}\right) }.$ \ Since 
$\tilde{\chi}\thickapprox \frac{1}{2M_{0}}\tilde{D}^{\dagger }\tilde{\phi}_{%
\tilde{n},\tilde{m}_{\tilde{s}}}\backsim \frac{1}{2M_{0}}\tilde{\beta}\tilde{
\phi}=\left( \frac{\tilde{\alpha}}{2M_{0}}\right) ^{3/4}\tilde{\phi}$, these
lower components vanish in the limit as $\tilde{\alpha}/2M_{0}\rightarrow 0$
. In this limit, solutions $\tilde{\psi}$ will have the same symmetry as the 
$\tilde{\phi},$ namely $SU\left( 4\right) \times SU\left( 2\right) .$

How do the $\tilde{\psi}_{\tilde{n},\tilde{m}_{\tilde{s}}}$ suggest quarks
and leptons? Consider the eigenfunctions of the $SU\left( 4\right) \;$%
\textbf{4} with, say, $SU\left( 2\right) $ spin ``up''. The \textbf{4}'s
upper components $\tilde{\phi}$ are $\tilde{x}^{1}\tilde{F}{\binom{1}{0}},\;%
\tilde{x}^{2}\tilde{F}{\binom{1}{0}},\;\tilde{x}^{3}\tilde{F}{\binom{1}{0}},$
and $\tilde{x}^{4}\tilde{F}{\binom{1}{0}},$ where $\tilde{F}\equiv \exp (-%
\frac{1}{2}\tilde{\beta}^{2}\tilde{X}^{2}).$ The triplet-subset $\left( 
\tilde{x}^{1},\;\tilde{x}^{2},\;\tilde{x}^{3}\right) \tilde{F}{\binom{1}{0}}$
, say, is an $SU\left( 3\right) $ \textbf{3} and can be coupled to another 
\textbf{3 }and an $SU\left( 3\right) $ \textbf{8 (}gluon) to yield an
invariant scalar, as we show in Sec. IV. Thus the\textbf{\ 3 }resembles a
color triplet. In addition, the \textbf{3} with spin ${\binom{1}{0}}$ and
the \textbf{3} from the other \textbf{4 }with spin ${\binom{0}{1}}$
constitute an $SU\left( 2\right) $\textbf{\ 2 }and can be coupled to an
isospin-\textbf{3} intermediate vector-boson to yield another invariant
scalar. Thus the \textbf{2} suggests a weak-isospin doublet. In this way the 
$SU\left( 3\right) \times SU\left( 2\right) $ $\left( \mathbf{3},\mathbf{2}
\right) $ resembles a quark.

The remaining wavefunction of the \textbf{4}, $\tilde{x}^{4}\tilde{F}{\binom{
1}{0}}$, will be an $SU\left( 4\right) $ ``color''-singlet, but still a
member of an $SU\left( 2\right) $-doublet. Thus it can be coupled to an
EW-boson, and suggests a lepton.

As noted earlier, weight diagrams of these multiplets are sketched in Fig.
2, for $\tilde{N}=0,\;1,$ and $2$. Dirac particles suggested by these
multiplets are indicated.

Dirac equation 11 predicts only two bound spin states for each ``orbital''
wavefunction $\tilde{\phi}_{\tilde{n}_{1}\tilde{n}_{2}\tilde{n}_{3}\tilde{n}%
_{4}}$, even though the Dirac wavefunction has four components. This is
analogous to the Dirac equation of the hydrogen atom which predicts just two
bound spin states for each orbital wavefunction rather than four; the other
two spin states correspond to positron-proton states and of course are not
bound. If the H-atom's confining potential were a scalar $\left(
-e^{2}/r\right) $ rather than the fourth component of a vector $\left(
-\gamma ^{0}e^{2}/r\right) $, then the positron-proton states would also be
bound, at a negative energy $E\thicksim -m$.

Similarly if the potential \textbf{V} in eight-dimensional equation 11 were $%
\tilde{V}$ rather than $\tilde{\gamma}^{5}\tilde{V},$ then a second spectrum
of confined states would appear with large \emph{lower} components $\tilde{%
\chi}\,\thickapprox \,\tilde{\phi}_{\tilde{n}_{1}\tilde{n}_{2}\tilde{n}_{3}%
\tilde{n}_{4}}\tilde{\xi}_{\tilde{m}_{\tilde{s}}}$ , small upper components $%
\tilde{\phi},$ and masses $M=-\overline{M}^{\left( \tilde{N}\right) }$. The
masses of this spectrum would be identical to those of the positive spectrum
since a negative mass can be interpreted as a positive mass with the
ordinary space-time spinor $\psi $ replaced by $\gamma ^{5}\psi $. This
would yield twice too many ``weak-isospin'' states. Thus the $\tilde{\gamma}%
^{5}\tilde{V}$ structure is indicated.

The potential $\tilde{V}$ generated by a soliton in five-dimensional $%
\mathbf{\phi }^{4}$-theory does not rise to infinity like a pure HO
potential, of course. Instead it reaches an upper limit, as shown in Fig. 1.
Similarly, any potential generated by a soliton in a higher dimensional
space might be expected to rise no higher than some maximum value, call it $%
\tilde{V}_{\infty }$. In the case of Dirac equation 11, if this maximum
height $\tilde{V}_{\infty }$ is $<<M_{0},$ then it is easy to show that the
equation generates confined states with masses $M$ in the range $%
M_{0}<M<M_{0}+V_{\infty }$ and free states (able to propagate anywhere in $%
M^{4}\times \tilde{R}^{4}$) with masses $M$ in the range $M_{0}+\tilde{V}
_{\infty }<M<\infty $. Because Eq. 11 is a relativistic equation, it also
admits free solutions with negative masses in the range $-\infty <M<-M_{0}+%
\tilde{V}_{\infty }$. Again, if the ordinary space-time spinor $\psi $ is
replaced by $\gamma ^{5}\psi ,$ then the masses in this third spectrum
change sign and span the range $M_{0}-\tilde{V}_{\infty }<M<\infty $.

Now this third mass-spectrum of free particles completely overlaps the
discrete spectrum of trapped particles, so is immediately ruled out by
experiment. However if all three spectra are lowered by an amount, say, $%
M_{0},$ so that $M$ is restricted to the three intervals $0<M<\tilde{V}
_{\infty }$, $\tilde{V}_{\infty }<M<\infty ,$ and $-\infty <M<-2M_{0}+\tilde{%
V}_{\infty },\;\;$then the third spectrum is equivalent to $2M_{0}-\tilde{V}
_{\infty }<M<$ $\infty $ and neither it nor the other continuous spectrum
overlaps the discrete spectrum.\thinspace \thinspace Furthermore this lowers
the threshold of discrete particles from $M_{0},$ which is far too massive,
to $0,$ closer to experiment.

Lowering these spectra can be accomplished, admittedly \textit{ad hoc},
simply by subtracting $M_{0}$ from $\tilde{V}$. Thus we will set 
\begin{equation}
\mathbf{V}=1\otimes \tilde{\gamma}^{5}\left( \tilde{V}-M_{0}\right) ,
\end{equation}
where $\tilde{V}$ is a HO-like potential that rises only to $\tilde{V}
_{\infty }$, analogous to the $\tilde{V}$ in Fig. 1. If lim$_{\tilde{X}
\rightarrow 0}\tilde{V}=\tilde{\alpha}^{3}\tilde{X}^{2}$ and $\left( 2\tilde{
N}+4\right) \tilde{\beta}^{2}/2M_{0}<<\tilde{V}_{\infty }$, then 
\begin{equation}
M\thickapprox \left( 2\tilde{N}+4\right) (\tilde{\beta}^{2}/2M_{0})=\left( 2%
\tilde{N}+4\right) (\tilde{\alpha}^{3}/2M_{0})^{1/2}\equiv M^{\left( \tilde{N%
}\right) },
\end{equation}
$\tilde{N}=0,\;1,\;2,\;.\;.\;.$

The possibility of continuous mass spectra raises an interesting question.
If we really are trapped in a topological or non-topological defect in a
higher-dimensional flat space, then are there particles freely ranging in
this extended space that could invade our local territory? Could they
sometimes appear even at energies above the Greisen-Zatsepin-Kus'min cutoff $%
[11]$ $(\thicksim 5\cdot 10^{19}$ eV)? And might we be able to propel Dirac
particles from our local space into that continuum? The threshold for
ejecting particles into the conjectured continuum may lie just above the
capability of present-day machines. In that case, there might be events such
as $e^{+}+e^{-}\rightarrow $ $2\;$free\ Dirac\ particles [1], appearing to
be $e^{+}+e^{-}\rightarrow $ nothing.

An interesting question is also posed by the discrete spectrum. If quarks
and leptons really do occur in $SU\left( 4\right) $ multiplets, then $%
SU\left( 3\right) $ \textbf{6}s should accompany the charm and strange
quarks in the $SU\left( 4\right) $ \textbf{10}s (see Fig. 2), and $SU\left(
3\right) $ \textbf{6}s and \textbf{10}s should accompany the top and bottom
quarks in $SU\left( 4\right) $ \textbf{20}s (not shown). Such particles have
never been seen$\footnote{%
Sextet Dirac particles have been conjectured for a long time. See, \textit{\
e.g.}, Ref. [12] and references cited therein.}$. If they do exist, then
they apparently do not interact with either the electroweak bosons or
gluons. This should make them good candidates for dark matter.

The discrete spectrum also suggests fourth-generation leptons, call them $%
\nu _{4}$ and $l_{4}.$ These particles would accompany the top and bottom
quarks in $SU\left( 4\right) $ \textbf{20}s. Such particles have never been
seen either. Present-day experimental limits [11] would place them above
about $45$ GeV.

Of course, just as in nuclear physics, not every member of a multiplet need
be bound. This might explain the absence of at least some of the \textbf{6}s
and \textbf{10}s and leptons.

Another interesting question has to do with the size of the bound states in
the extra dimensions. Their widths can be easily calculated if the confining
``potential'' is taken to be a harmonic oscillator. If we adopt as a measure
of this width (squared)

\begin{equation}
\int \tilde{\psi}_{\widetilde{n}}^{\dagger }\tilde{X}^{2}\tilde{\psi}_{%
\widetilde{n}}d^{4}\tilde{x}\approx \int \tilde{\phi}_{\widetilde{n}%
}^{\dagger }\tilde{X}^{2}\tilde{\phi}_{\widetilde{n}}d^{4}\tilde{x}\equiv 
\tilde{X}_{rms}^{2}
\end{equation}
[recall that $\widetilde{X}^{2}=\sum_{\widetilde{\mu }=1}^{4}\left( 
\widetilde{x}^{\widetilde{\mu }}\right) ^{2}$]$,$ then one can readily show
that $\tilde{X}_{rms}^{2}=\left( \tilde{N}+2\right) /\tilde{\beta}^{2}.$
Folding in Eq. 26 and restoring $\mathbf{\rlap/}h$ and $c$ yield 
\begin{equation}
\tilde{X}_{\widetilde{N}}=\left( \tilde{N}+2\right) \left( M^{\left( 
\widetilde{N}\right) }M_{0}\right) ^{-\frac{1}{2}}\mathbf{\rlap/}h/c\equiv 
\tilde{X}_{rms}.
\end{equation}
As for numbers, if we choose $M_{0}=100\;$TeV and adjust $\tilde{\alpha}$ to
fit an up-quark mass of $6$ MeV, then $\tilde{X}_{rms}=0.024$\ F (with $%
\tilde{\alpha}=585$ MeV). If we set $M_{0}=10^{4}\;$TeV, then $\tilde{X}%
_{rms}=0.0024$\ F (with $\tilde{\alpha}=2710$ MeV).

Now according to the Particle Data Group [11], leptons and quarks have radii
no greater than the order of $10^{-4}$ F. However these are radii in \emph{%
ordinary} space-time. They can be measured, \textit{e.g.}, by probing the
Dirac particle with photon-wavefunctions $\exp \left( -ip\cdot x\right) $
over a range of momenta $\overrightarrow{p}$ and taking the Fourier
transform. To measure the size of a Dirac particle in both ordinary space
and extended space, one could figuratively probe the particle with
``photon'' wavefunctions $\exp \left( -ip\cdot x+i\tilde{p}\cdot \tilde{x}
\right) $ where now the higher dimensional ``momentum'' $\tilde{p}$ would
also range over some interval. However these would be the wavefunctions of
probes free to propagate in all of the dimensions, higher and lower. Even if
such particles were to exist, we would only have access to the photons
confined in $M^{4}.$ These ordinary photons have just a single wavefunction
in the higher dimensions (more kinds of wavefunctions would mean more kinds
of photons), so they always weight the wavefunction of the Dirac particle
the same way in the higher dimensions and tell us nothing about its
higher-dimensional structure. If this is typical of all types of
higher-dimensional reactions measurable by us, then the size of the Dirac
particles in the extra dimensions is not subject to the Particle Data Group
limits.

\section{Interaction of Dirac particles with gauge fields}

\subsection{QCD interaction}

In this section we will indicate how the ``quark'' solutions of Eq. 11 might
be combined with a ``gluon'' gauge field to reproduce the QCD interaction
lagrangian in the limit as $V_{\infty }\rightarrow \infty $. Consider the
eight-dimensional interaction lagrangian 
\begin{equation}
\mathcal{L}_{int}=\tfrac{1}{\sqrt{2}}g_{s}\sum_{\mu =0}^{3}\sum_{\tilde{i},%
\tilde{j}=1}^{3}\overline{\mathbf{\psi }}\gamma ^{\mu }\tilde{E}_{\tilde{i}%
\tilde{j}}\mathbf{G}_{\mu }^{\tilde{i}\tilde{j}}\mathbf{\psi ,}
\end{equation}
where $\mathbf{G}_{\mu }^{\tilde{i}\tilde{j}}$ is the ``gauge field'' and
the $\tilde{E}_{\tilde{i}\tilde{j}}$ are the $SU\left( 3\right) $-generators
defined by Eq. 47 in Appendix A. It may not be unreasonable to suppose that $%
\mathbf{G}_{\mu }^{\tilde{i}\tilde{j}}=G_{\mu }(x)\tilde{G}^{\tilde{i}\tilde{
j}}(\tilde{x})$, in which case ordinary space-time quantities such as
energy-momentum, angular momentum, \textit{etc}. are conserved. In this
case, $\mathcal{L}_{int}$ will be a sum over terms of the sort 
\begin{equation}
\tfrac{1}{\sqrt{2}}g_{s}\overline{\psi }_{f}\left( x\right) \gamma ^{\mu
}G_{s,\mu }\left( x\right) \psi _{i}\left( x\right) \cdot \overline{\tilde{
\psi}}_{\tilde{f}}(\tilde{x})\tilde{E}_{\tilde{i}\tilde{j}}\tilde{G}_{\tilde{%
s}}^{\tilde{i}\tilde{j}}(\tilde{x})\tilde{\psi}_{\tilde{i}}(\tilde{x}),
\end{equation}
where Dirac and Bose states are identified by $f$, $i$, and $s$ in $M^{4}$
and $\tilde{f}$, $\tilde{i}$, and $\tilde{s}$ in $\tilde{R}^{4}$. ($%
\overline{\tilde{\psi}}=\tilde{\psi}^{\dagger }\tilde{\gamma}^{5}.)$ In the
case of the bound-state solutions $\tilde{\psi}_{\widetilde{n}_{1}\widetilde{%
n}_{2}\widetilde{n}_{3}\widetilde{n}_{4}\widetilde{m}_{\tilde{s}}},$ the
field $\tilde{G}_{\tilde{s}}^{\tilde{i}\tilde{j}}(\tilde{x})$ can cause
transitions to different ``families'' ($\widetilde{n}_{4})$, to different $%
SU\left( 3\right) $-multiplets $(\widetilde{N}^{\prime }=\widetilde{n}_{1}+%
\widetilde{n}_{2}+\widetilde{n}_{3})$ (\emph{e.g., }quark$\rightarrow $
lepton), and to the opposite ``weak-isospin'' state ($\widetilde{m}_{\tilde{s%
}}$), all of which are inconsistent with QCD. However if $\tilde{G}_{\tilde{s%
}}^{\tilde{i}\tilde{j}}(\tilde{x})$ is essentially constant over the range
of integration of $\overline{\tilde{\psi}}_{\tilde{f}}(\tilde{x})\tilde{E}_{%
\tilde{i}\tilde{j}}\tilde{G}_{\tilde{s}}^{\tilde{i}\tilde{j}}(\tilde{x})%
\tilde{\psi}_{\tilde{i}}(\tilde{x})$ in $\tilde{R}^{4}$, then none of these
transitions can take place due to the orthogonality of the
harmonic-oscillator functions, at least in the limit $\tilde{\alpha}
/2M_{0}\rightarrow 0$ where $\tilde{\psi}\rightarrow \left( \tilde{\phi}_{%
\widetilde{n}_{1}\widetilde{n}_{2}\widetilde{n}_{3}\widetilde{n}_{4,}%
\widetilde{m}_{\tilde{s}}}(\tilde{x}),\;0\right) ^{T}$. An example of a
field that would be essentially constant over the range of integration is $%
\tilde{G}_{\tilde{s}}^{\tilde{i}\tilde{j}}(\tilde{x})=\widehat{G}_{\tilde{s}%
}^{\tilde{i}\tilde{j}}\exp \left( -\frac{1}{2}\tilde{\gamma}^{2}\tilde{X}
^{2}\right) $, where the range $\tilde{\gamma}^{-1}$ of the gluon field is
much greater than the range $\tilde{\beta}^{-1}$ of the Dirac-particle
fields $\left( \tilde{\gamma}^{-1}>>\tilde{\beta}^{-1}\right) $; $\widehat{G}
_{\tilde{s}}^{\tilde{i}\tilde{j}}$ is a constant which identifies the
particular gluon field $\tilde{s}$.

Lagrangian 30 is still not consistent with QCD since $\tilde{\psi}$ can be
any $SU\left( 3\right) $-multiplet: \textbf{1}, \textbf{3}, \textbf{6}, 
\textbf{10}, . . . However if we limit the $SU\left( 3\right) $-states to
just \textbf{3}s ($\widetilde{n}_{1}+\widetilde{n}_{2}+\widetilde{n}_{3}=1$
``quarks'') and \textbf{1}s ($\widetilde{n}_{1}+\widetilde{n}_{2}+\widetilde{%
n}_{3}=0$ ``leptons''), then the lagrangian is equivalent to the QCD
lagrangian\footnote{%
If by no other means, one can project out \textbf{6}s (at least at tree
level) by inserting the operator $\left( 2-\sum_{\tilde{k}=1}^{3}\widetilde{a%
}_{\tilde{k}}^{\dagger }\widetilde{a}_{\tilde{k}}\right) $ before or after $%
\tilde{E}_{\tilde{i}\tilde{j}}$ in lagrangian 29. Higher multiplets, if they
are exist (\emph{i.e.}, are bound), can be projected out by similar
operators. \textbf{1}s make no contribution since $\tilde{E}_{\tilde{i}%
\tilde{j}}\tilde{\phi}_{000}=0.$}, as we will now show. One can integrate
over $\tilde{R}^{4}$ and replace the $\tilde{\phi}_{\widetilde{n}_{1}%
\widetilde{n}_{2}\widetilde{n}_{3}\widetilde{n}_{4,}\widetilde{m}_{\tilde{s}%
}}(\tilde{x})$ by unit column vectors $\widehat{\phi }_{\widetilde{n}_{1}%
\widetilde{n}_{2}\widetilde{n}_{3}\widetilde{n}_{4,}\widetilde{m}_{\tilde{s}%
}}=\widehat{\phi }_{\widetilde{n}_{1}\widetilde{n}_{2}\widetilde{n}%
_{3}}\otimes \widehat{\phi }_{\widetilde{n}_{4}}\otimes \tilde{\xi}_{\tilde{m%
}_{\tilde{s}}}$, where the $\widehat{\phi }_{\widetilde{n}_{1}\widetilde{n}%
_{2}\widetilde{n}_{3}}\equiv \widehat{\phi }_{color}$ comprise the ``color''
states [ $\widehat{\phi }_{100}=(1,0,0)^{T},$ $\widehat{\phi }%
_{010}=(0,1,0)^{T},$ and $\widehat{\phi }_{001}=(0,0,1)^{T}$], $\widehat{%
\phi }_{\widetilde{n}_{4}}$ is a column vector with an element for each
``generation'', and $\tilde{\xi}_{\tilde{m}_{\tilde{s}}}=(1,0)^{T}$ or $%
(0,1)^{T}$, the previously defined ``weak-isospin'' spinor. The operators $%
\tilde{E}_{\tilde{i}\tilde{j}}(\tilde{x},$ $\partial \tilde{x})$ can
similarly be replaced by $3\times 3$ matrices $\widehat{E}_{\tilde{i}\tilde{j%
}},$ where

\begin{equation}
\widehat{E}_{11}=\left( 
\begin{array}{ccc}
\frac{2}{3} & 0 & 0 \\ 
0 & -\frac{1}{3} & 0 \\ 
0 & 0 & -\frac{1}{3}
\end{array}
\right) ,\;\widehat{E}_{12}=\left( 
\begin{array}{lll}
0 & 1 & 0 \\ 
0 & 0 & 0 \\ 
0 & 0 & 0
\end{array}
\right) ,\;\widehat{E}_{13}=\left( 
\begin{array}{lll}
0 & 0 & 1 \\ 
0 & 0 & 0 \\ 
0 & 0 & 0
\end{array}
\right)
\end{equation}
and similarly for $\widehat{E}_{2\tilde{j}}$ and $\widehat{E}_{3\tilde{j}},\;%
\tilde{j}=1,\;2,\;3$. The interaction lagrangian now takes the form 
\begin{equation}
\widehat{\mathcal{L}}_{int}=\tfrac{1}{\sqrt{2}}g_{s}\sum_{\mu =0}^{3}\sum_{%
\tilde{i},\tilde{j}=1}^{3}\overline{\widehat{\mathbf{\psi }}}\gamma ^{\mu }%
\widehat{E}_{\tilde{i}\tilde{j}}\widehat{\mathbf{G}}_{\mu }^{\tilde{i}\tilde{
j}}\widehat{\mathbf{\psi }},
\end{equation}
where $\widehat{\mathbf{\psi }}$ is a sum over states $\psi \left( x\right)
\otimes \widehat{\phi }_{color}\otimes \widehat{\phi }_{\widetilde{n}
_{4}}\otimes \tilde{\xi}_{\tilde{m}_{\tilde{s}}}$ and $\widehat{\mathbf{G}}%
_{\mu }^{\tilde{i}\tilde{j}}$ is a sum over states $G_{\mu }\left( x\right) 
\widehat{G}^{\tilde{i}\tilde{j}}$. This lagrangian equals the QCD lagrangian
[11] 
\begin{equation}
\mathcal{L}_{int}^{\scriptstyle QCD}=\tfrac{1}{2}g_{s}\sum_{\mu
=0}^{3}\sum_{a=1}^{8}\overline{\widehat{\mathbf{\psi }}}\gamma ^{\mu
}\lambda ^{a}\mathbf{G}_{\mu }^{a}\widehat{\mathbf{\psi }},
\end{equation}
where the $\lambda ^{a}$ are the Gell-Mann $3\times 3\;SU\left( 3\right) $
matrices, if the QCD gauge fields 
\[
\mathbf{G}_{\mu }^{1}=\tfrac{1}{\sqrt{2}}\left( \widehat{\mathbf{G}}_{\mu
}^{12}+\widehat{\mathbf{G}}_{\mu }^{21}\right) ,\ \;\mathbf{G}_{\mu }^{2}=%
\tfrac{i}{\sqrt{2}}\left( \widehat{\mathbf{G}}_{\mu }^{12}-\widehat{\mathbf{G%
}}_{\mu }^{21}\right) ,\;\; 
\]
\[
\mathbf{G}_{\mu }^{3}=\tfrac{1}{\sqrt{2}}\left( \widehat{\mathbf{G}}_{\mu
}^{11}-\widehat{\mathbf{G}}_{\mu }^{22}\right) ,\;\;\mathbf{G}_{\mu }^{4}=%
\tfrac{1}{\sqrt{2}}\left( \widehat{\mathbf{G}}_{\mu }^{13}+\widehat{\mathbf{G%
}}_{\mu }^{31}\right) ,\;\; 
\]
\[
\mathbf{G}_{\mu }^{5}=\tfrac{i}{\sqrt{2}}\left( \widehat{\mathbf{G}}_{\mu
}^{13}-\widehat{\mathbf{G}}_{\mu }^{31}\right) ,\;\;\mathbf{G}_{\mu }^{6}=%
\tfrac{1}{\sqrt{2}}\left( \widehat{\mathbf{G}}_{\mu }^{23}+\widehat{\mathbf{G%
}}_{\mu }^{32}\right) ,\;\; 
\]
\begin{equation}
\mathbf{G}_{\mu }^{7}=\tfrac{i}{\sqrt{2}}\left( \widehat{\mathbf{G}}_{\mu
}^{23}-\widehat{\mathbf{G}}_{\mu }^{32}\right) ,\ \ \mathbf{G}_{\mu }^{8}=%
\tfrac{1}{\sqrt{6}}\left( \widehat{\mathbf{G}}_{\mu }^{11}+\widehat{\mathbf{G%
}}_{\mu }^{22}-2\widehat{\mathbf{G}}_{\mu }^{33}\right) .\bigskip
\end{equation}

\subsection{Electroweak interaction}

We can also construct a lagrangian which reduces to the electroweak
interaction, provided the latter is of the ``left-right symmetric'' variety
[7]. Consider the eight-dimensional lagrangian 
\[
\mathcal{L}_{int}=\tfrac{1}{2}\sum_{\mu =0}^{3}\sum_{\tilde{j}=1}^{3}\left[
g_{L}\overline{\mathbf{\psi }}\gamma ^{\mu }\tfrac{1}{2}\left( 1-\gamma
^{5}\right) \tilde{\Sigma }^{\tilde{j}}\mathbf{W}_{L,\mu }^{\tilde{j}}%
\mathbf{\psi }\right. \mathbf{\,}
\]
\begin{equation}
\left. \mathbf{+\,}~g_{R}\overline{\mathbf{\psi }}\gamma ^{\mu }\tfrac{1}{2}%
\left( 1+\gamma ^{5}\right) \tilde{\Sigma }^{\tilde{j}}\mathbf{W}_{R,\mu }^{%
\tilde{j}}\mathbf{\psi }\right] \mathbf{,}
\end{equation}
where $\mathbf{W}_{L,\mu }^{\tilde{j}}$ and $\mathbf{W}_{R,\mu }^{\tilde{j}}$
are ``electro-weak'' gauge-field triplets and $\tilde{\Sigma }^{\tilde{j}%
}=\left( 
\begin{array}{ll}
\tilde{\sigma}^{\tilde{j}} & 0 \\ 
0 & \tilde{\sigma}^{\tilde{j}}
\end{array}
\right) ,$ $\tilde{j}=1,\;2,\;3$. If we again assume that gauge fields can
be factored, \textit{i.e.}, $\mathbf{W}_{L,\mu }^{\tilde{j}}=W_{L,\mu
}\left( x\right) \tilde{W}^{\tilde{j}}(\tilde{x})$ and $\mathbf{W}_{R,\mu }^{%
\tilde{j}}=W_{R,\mu }\left( x\right) \tilde{W}^{\tilde{j}}(\tilde{x}),$ then
Minkowski--space quantities will be conserved. If in addition we
parameterize $\tilde{W}^{\tilde{j}}(\tilde{x})=\widehat{W}^{\tilde{j}}\exp
\left( -\frac{1}{2}\tilde{\gamma}^{2}\tilde{X}^{2}\right) $ where $\widehat{W%
}^{\tilde{j}}$ is a constant identifying the boson, and again assume that $%
\tilde{\gamma}^{-1}>>\tilde{\beta}^{-1}$, then in the limit $\tilde{\alpha}%
/2M_{0}\rightarrow 0$ where orthogonality holds, Dirac particles may not
make transitions to different ``generations'', to different $SU\left(
3\right) $-multiplets, or to different states within an $SU\left( 3\right) $
-multiplet. However transitions to opposite ``weak-isospin'' states may take
place due to the $\tilde{\sigma}^{\tilde{j}}$-operator. If again we limit
Dirac states to $SU\left( 3\right) $ \textbf{1}s and \textbf{3}s, then
integrating lagrangian 35 over $\tilde{R}^{4}$ yields 
\[
\widehat{\mathcal{L}}_{int}=\tfrac{1}{2}\sum_{\mu =0}^{3}\sum_{\tilde{j}%
=1}^{3}\left[ g_{L}\overline{\widehat{\mathbf{\psi }}}\gamma ^{\mu }\tfrac{1%
}{2}\left( 1-\gamma ^{5}\right) W_{L,\mu }\left( x\right) \widetilde{\sigma }%
^{\tilde{j}}\widehat{W}^{\tilde{j}}\widehat{\mathbf{\psi }}\right. 
\]
\begin{equation}
\left. \mathbf{+\,\,}g_{R}\overline{\widehat{\mathbf{\psi }}}\gamma ^{\mu }%
\tfrac{1}{2}\left( 1+\gamma ^{5}\right) W_{R,\mu }\left( x\right) \widetilde{%
\sigma }^{\tilde{j}}\widehat{W}^{\tilde{j}}\widehat{\mathbf{\psi }}\right] ,
\end{equation}
where operator $\widehat{\mathbf{\psi }}$ is a sum over states $\psi \left(
x\right) \otimes \widehat{\phi }_{\widetilde{n}_{1}\widetilde{n}_{2}%
\widetilde{n}_{3}}\otimes \widehat{\phi }_{\widetilde{n}_{4}}\otimes \tilde{%
\xi}_{\widetilde{m}_{\tilde{s}}}$with $\widetilde{n}_{1}+\widetilde{n}_{2}+%
\widetilde{n}_{3}=0$ or $1.$ If we denote $\overrightarrow{\tilde{\sigma}}%
\equiv \overrightarrow{\tau }$, $W_{L,\mu }\left( x\right) \widehat{W}^{%
\tilde{j}}\equiv \mathbf{W}_{L,\mu }^{\tilde{j}}\left( x\right) ,$ and $%
W_{R,\mu }\left( x\right) \widehat{W}^{\tilde{j}}\equiv \mathbf{W}_{R,\mu }^{%
\tilde{j}}\left( x\right) ,\;\tilde{j}=1,\;2,\;3,$ then lagrangian 36 takes
the form\pagebreak 
\[
\widehat{\mathcal{L}}_{int}=\tfrac{1}{2}g_{L}\overline{\widehat{\mathbf{\psi 
}}}\gamma ^{\mu }\tfrac{1}{2}\left( 1-\gamma ^{5}\right) \overrightarrow{%
\tau }\cdot \overrightarrow{\mathbf{W}}_{L,\mu }\left( x\right) \widehat{%
\mathbf{\psi }}\mathbf{\;}
\]
\begin{equation}
\mathbf{+\;}\tfrac{1}{2}g_{R}\overline{\widehat{\mathbf{\psi }}}\gamma ^{\mu
}\tfrac{1}{2}\left( 1+\gamma ^{5}\right) \overrightarrow{\tau }\cdot 
\overrightarrow{\mathbf{W}}_{R,\mu }\left( x\right) \widehat{\mathbf{\psi }}.
\end{equation}
This lagrangian forbids family-changing transitions, quark$\leftrightarrow $
lepton transitions, and, in the case of quarks, color-changing transitions.
If a fourth, weak-isospin-zero gauge field $B^{\mu }$ is introduced with
appropriate, distinct couplings to quark and lepton fields, and CKM mixing
is accommodated, then the interaction lagrangian becomes a
left-right-symmetric version of the EW interaction in the Standard Model.

\section{Discussion}

We have considered the possibility that quarks and leptons live in a flat
higher dimensional space, confined to a small volume in the extra
dimensions, perhaps by a soliton. If the confinement is due to a
harmonic-oscillator ``potential'' which only rises to a certain height, plus
a large constant $M_{0}$, then four extra dimensions suffice to generate
harmonic-oscillator Dirac-particle states which bear some resemblance to the
quarks and leptons of the Standard Model. The symmetry group of the states
in the higher dimensions is just $SO\left( 4\right) $, but thanks to the
degenerate levels of the harmonic-oscillator potential, the $SO\left(
4\right) $ symmetry is enlarged to an approximate $SU\left( 4\right) $
symmetry. Furthermore, because the ``spin'' states are decoupled from the
``orbital'' states, an additional approximate $SU\left( 2\right) $ symmetry
occurs in these same four extra dimensions. Thus the HO potential generates
an approximate $SU\left( 4\right) \times SU\left( 2\right) $ symmetry.

In an $SU\left( 4\right) =SU\left( 3\right) \times U\left( 1\right) $
decomposition, the $SU\left( 3\right) $ representations \textbf{1, 3}, 
\textbf{6}, . . resemble leptons, quarks, and new kinds of Dirac particles,
respectively; the $U\left( 1\right) $ representations introduce replications
which might be identified with generations. The $SU\left( 2\right) $
symmetry is represented by doublets reduced from Dirac spinors and is
similar to weak isospin.

We have shown how the Dirac particles might be coupled to ``gluons'' to
reproduce the QCD interaction, and to ``EW bosons'' to reproduce a
left-right symmetric extension of the electroweak interaction. Because the $%
SU\left( 4\right) $ symmetry is realized in just four dimensions instead of
eight, only real solutions corresponding to tetrahedral weight diagrams are
generated. But these are the only multiplets needed to represent the quarks
and leptons of the Standard Model.

The $SU\left( 4\right) $ symmetry is not exact, primarily because the
harmonic-oscillator well rises only a finite amount. On the other hand, this
finite rise limits the number of quarks and leptons in a natural way. (The
symmetry in the lower generations would become more exact should the well
rise higher to accommodate additional generations.)

Also the masses predicted by the harmonic-oscillator well rise only linearly
with generation number, whereas in nature the masses seem to rise
exponentially. However the particles' masses might be augmented by a
Higgs-Yukawa mechanism as in the Standard Model.

If a simple potential in four flat extra dimensions can yield Dirac bound
states similar in many ways to those of quarks and leptons, then it may be
worthwhile to search for an underlying physical mechanism, perhaps an
instanton arising from a generalization of the Rubakov-\-Shaposhnikov [1] or
Kaplan [2] lagrangian to eight dimensions. Hopefully such a lagrangian will
not generate the $SU\left( 3\right) $ \textbf{6}s and \textbf{10}s which
occur in the higher $SU\left( 4\right) $ multiplets predicted by our present
model. On the other hand, if the \textbf{6}s and \textbf{10}s were decoupled
from the gauge bosons, then they would be interesting candidates for dark
matter.

\section*{Acknowledgments}

I would like to thank George Sudarshan, Dick Arnowitt, Chris Pope, Mike
Duff, Ergin Sezgin and Rabindra Mohapatra for helpful discussions.

\section*{Note added in proof}

Recently there has been considerable interest in the possibility that the
extra dimensions, while not flat or infinite in extent, may still be of
millimeter size. Akrani-Hamed, Dimopoulos, Dvali, and Antoniadis and others
have proposed models of this sort. See, \emph{e. g.}, references [14] --
[16]. These models have passed a gauntlet of tests, but many remain. See, 
\emph{e. g.}, reference [17]. 

\section*{References}

1. V.~A.~Rubakov and M.~E.~Shaposhnikov, ``Do we live inside a domain
wall?'', Phys. Lett. B \textbf{125,} 136 (1983).

2. D. B. Kaplan, ``A method for simulating chiral fermions on the lattice'',
Phys. Lett. B \textbf{288,} 342 (1992).

3. H. B. Nielsen and M. Ninomiya, ``A no-go theorem for regularizing chiral
fermions'', Phys. Lett. B \textbf{105,} 219 (1981); L. H. Karsten and
J.~Smit, ``Lattice fermions: species doubling, chiral invariance and the
triangle anomaly'', Nucl. Phys. B \textbf{183,} 103 (1981).

4. K. Jansen, ``Domain wall fermions and chiral gauge theories'', Phys. Rep. 
\textbf{273, }1\textbf{\ }(1996).

5. R.~A.~Bryan, ``Isoscalar quarks and leptons in an eight-dimensional
space'', Phys. Rev. D \textbf{34,} 1184 (1986).

6. G. Dvali and M. Shifman, ``Domain walls in strongly coupled theories'',
Phys. Lett. B \textbf{396,} 64 (1997); hep-th/9612128.

7. \textit{e.g.}, R. N. Mohapatra and G. Senjanovi\'{c}, ``Neutrino masses
and mixings in gauge models with spontaneous parity violation'',
Phys.~Rev.~D \textbf{23,} 165 (1981).

8. R. Rajaraman, \emph{Solitons and instantons} (Amsterdam, North-Holland,
1982) p.~140.

9. G. Dvali and M. Shifman, ``Dynamical compactification as a mechanism of
spontaneous supersymmetry breaking'', Nucl$.$ Phys. B \textbf{504,} 127
(1997); hep-th/9611213.

10. \textit{e.g.}, R. A. Bryan and M. M. Davenport, ``Spinorial
representations of SU(3) from a factored harmonic-oscillator equation'', J.
Phys. A \textbf{29,} 3129 (1996).

11. R. M. Barnett \emph{et al.} (Particle Data Group), ``Review of particle
physics'', Phys. Rev. D \textbf{54,} 1 (1996); K. Greisen, Phys. Rev. Lett. 
\textbf{16}, 748 (1966); G. T. Zatsepin and V. A. Kuz'min, Pis'ma Zh.
\'{E}ksp. Teor. Fiz. \textbf{4}, 114 (1966) [JETP Lett. \textbf{4}, 78
(1966].

12. P. H. Frampton and S. L. Glashow, ``Unifiable chiral color with natural
Glashow-Iliopoulos-Maiani mechanism'', Phys. Rev. Letters \textbf{58,} 2168
(1987); T. E. Clark, C. N. Leung, S. T. Love, and J. L. Rosner, ``Sextet
quarks and light pseudoscalars'', Phys. Letters B \textbf{177,} 413 (1986).

13. H.~J.~Lipkin, \emph{Lie Groups for Pedestrians, 2nd edn} (Amsterdam,
North-Holland, 1966) p.~33.

14. N. Arkani--Hamed, S. Dimopoulos, and G. Dvali, ``The hierarchy problem
and new dimensions at a millimeter'', Physics Letters B \textbf{429}, 263
(1998); hep-ph/9803315.

15. I. Antoniadis, N. Arkani--Hamed, S. Dimopoulos, and G. Dvali, ``New
dimensions at a millimeter to a fermi and superstrings at a TeV'', Physics
Letters B \textbf{436}, 257 (1998); hep-ph/9804398.

16. N. Arkani--Hamed, S. Dimopoulos, and G. Dvali, ``Phenomenology,
astrophysics and cosmology of theories with sub-millimeter dimensions and
TeV scale quantum gravity'', Physical Review D \textbf{59}, 086004 (1999);
hep-ph/9807344.

17. T. Banks, M. Dine, and A. E. Nelson, ``Constraints on theories with
large extra dimensions'', hep-th/9903019 v3.

\section*{Appendix A: Harmonic-oscillator symmetry groups}

Consider the differential equation 
\begin{equation}
\left( \widetilde{\square }-\tilde{\beta}^{4}\tilde{X}^{2}+m^{2}\right) 
\widetilde{\phi }=0
\end{equation}
in four Euclidean extra dimensions $\tilde{R}^{4}\left( \tilde{x}^{1},\;%
\tilde{x}^{2},\;\tilde{x}^{3},\;\tilde{x}^{4}\right) ,$ with $\tilde{\beta}$
a real constant, $\widetilde{\square }$ $=\sum_{\widetilde{\mu }
=1}^{4}\partial ^{2}/\partial \left( \widetilde{x}^{\tilde{\mu}}\right) ^{2}$
and $\tilde{X}^{2}$ $=\sum_{\widetilde{\mu }=1}^{4}\left( \widetilde{x}^{%
\tilde{\mu}}\right) ^{2}$. This equation has (unnormalized) solutions 
\begin{equation}
\widetilde{\phi }=\widetilde{\phi }_{\widetilde{n}_{1}}\left( \tilde{\beta}%
\widetilde{x}^{1}\right) \widetilde{\phi }_{\widetilde{n}_{2}}\left( \tilde{
\beta}\widetilde{x}^{2}\right) \widetilde{\phi }_{\widetilde{n}_{3}}\left( 
\tilde{\beta}\widetilde{x}^{3}\right) \widetilde{\phi }_{\widetilde{n}%
_{4}}\left( \tilde{\beta}\widetilde{x}^{4}\right) \equiv \widetilde{\phi }_{%
\widetilde{n}_{1}\widetilde{n}_{2}\widetilde{n}_{3}\widetilde{n}_{4}}
\end{equation}
with eigenvalues 
\begin{equation}
m^{2}=\left( 2\widetilde{N}+4\right) \tilde{\beta}^{2},
\end{equation}
where 
\begin{equation}
\widetilde{\phi }_{\widetilde{n}_{\widetilde{\mu }}}\left( \tilde{\beta}%
\widetilde{x}^{\widetilde{\mu }}\right) =H_{\widetilde{n}_{\widetilde{\mu }%
}}\left( \widetilde{\beta }\widetilde{x}^{\widetilde{\mu }}\right) \exp
\left[ -\tfrac{1}{2}\tilde{\beta}^{2}(\tilde{x}^{\tilde{\mu}})^{2}\right]
,\;\;\;\;\widetilde{\mu }=1,\ 2,\ 3,\ 4.
\end{equation}
Here $H_{\widetilde{n}_{\widetilde{\mu }}}$ is a Hermite polynomial of
degree $\widetilde{n}_{\widetilde{\mu }}=0,$\ $1,\;2,\;.\;.$, and$\;%
\widetilde{N}=\sum_{\widetilde{\mu }=1}^{4}\widetilde{n}_{\tilde{\mu}}=0,\
1,\;2,\;.\;$.

Solutions of a given $\widetilde{N}\,\,\,$form multiplets of common mass and
can be transmuted one into another by the generators 
\begin{equation}
\widetilde{a}_{\tilde{\mu}}^{\dagger }\widetilde{a}_{\tilde{\nu}}-\tfrac{1}{4%
}\delta _{\tilde{\mu}\tilde{\nu}}\sum_{\widetilde{\rho }=1}^{4}\widetilde{a}%
_{\tilde{\rho}}^{\dagger }\widetilde{a}_{\tilde{\rho}}\equiv \tilde{E}_{%
\tilde{\mu}\tilde{\nu}},\;\;\;\mu ,\;\nu =1,\;2,\;3,\;4,
\end{equation}
where 
\begin{equation}
\widetilde{a}_{\tilde{\mu}}^{\dagger }=\tfrac{1}{\sqrt{2}}\left( \widetilde{
\beta }\widetilde{x}^{\widetilde{\mu }}-\widetilde{\beta }^{-1}\partial
/\partial \widetilde{x}^{\widetilde{\mu }}\right)
\end{equation}
and 
\begin{equation}
\widetilde{a}_{\tilde{\mu}}=\tfrac{1}{\sqrt{2}}\left( \widetilde{\beta }%
\widetilde{x}^{\widetilde{\mu }}+\widetilde{\beta }^{-1}\partial /\partial 
\widetilde{x}^{\widetilde{\mu }}\right) .
\end{equation}
Operating on the $\widetilde{\phi }_{\widetilde{n}_{1}\widetilde{n}_{2}%
\widetilde{n}_{3}\widetilde{n}_{4}}$, the $\tilde{E}_{\tilde{\mu}\tilde{\nu}
} $ satisfy 
\begin{equation}
\left[ \tilde{E}_{\tilde{\mu}\tilde{\nu}},\;\tilde{E}_{\tilde{\rho}\tilde{%
\sigma}}\right] =\delta _{\widetilde{\nu }\widetilde{\rho }}\tilde{E}_{%
\tilde{\mu}\tilde{\sigma}}-\delta _{\widetilde{\sigma }\widetilde{\mu }}%
\tilde{E}_{\tilde{\rho}\tilde{\nu}},\;\;\;\tilde{\mu},\;\tilde{\nu},\;\tilde{
\rho},\;\tilde{\sigma}=1,\;2,\;3,\;4.
\end{equation}
But this is the vector-multiplication rule of $SU\left( 4\right) $.
Therefore the fifteen independent $\tilde{E}_{\tilde{\mu}\tilde{\nu}}$
comprise an $SU\left( 4\right) $-algebra and the $\widetilde{\phi }_{%
\widetilde{n}_{1}\widetilde{n}_{2}\widetilde{n}_{3}\widetilde{n}_{4}}$ are
its representations [5, 13]. The smallest representation, with $\widetilde{N}
=0$, is the \textbf{1} consisting of $\;\tilde{\phi}_{0000}=\exp \left( -%
\frac{1}{2}\tilde{\beta}^{2}\tilde{X}^{2}\right) \equiv \tilde{F},$ where $%
\widetilde{X}^{2}=\sum_{\widetilde{\mu }=1}^{4}\left( \widetilde{x}^{%
\widetilde{\mu }}\right) ^{2}$. The next smallest, with $\widetilde{N}=1,$
is the \textbf{4} consisting of\ $\tilde{\phi}_{1000}=\tilde{x}^{1}\tilde{F}
,\;\tilde{\phi}_{0100}=\tilde{x}^{2}\tilde{F},\;\tilde{\phi}_{0010}=\tilde{x}
^{3}\tilde{F},\;$and $\tilde{\phi}_{0001}=\tilde{x}^{4}\tilde{F}.$

Likewise the solutions 
\begin{equation}
\widetilde{\phi }_{\widetilde{n}_{1}}\left( \tilde{\beta}\widetilde{x}%
^{1}\right) \widetilde{\phi }_{\widetilde{n}_{2}}\left( \tilde{\beta}%
\widetilde{x}^{2}\right) \widetilde{\phi }_{\widetilde{n}_{3}}\left( \tilde{%
\beta}\widetilde{x}^{3}\right) \equiv \widetilde{\phi }_{\widetilde{n}_{1}%
\widetilde{n}_{2}\widetilde{n}_{3}}
\end{equation}
are basis functions of an $SU\left( 3\right) $-algebra 
\begin{equation}
\tilde{E}_{\tilde{i}\tilde{j}}=\widetilde{a}_{\tilde{i}}^{\dagger }%
\widetilde{a}_{\tilde{j}}-\tfrac{1}{3}\delta _{\tilde{i}\tilde{j}}\sum_{%
\widetilde{k}=1}^{3}\widetilde{a}_{\tilde{k}}^{\dagger }\widetilde{a}_{%
\tilde{k}},\;\;\;\tilde{i},\tilde{j}=1,\;2,\;3,
\end{equation}
where $\widetilde{a}_{\tilde{i}}^{\dagger }$ and $\widetilde{a}_{\tilde{j}}$
are defined like $\widetilde{a}_{\tilde{\mu}}^{\dagger }$ and $\widetilde{a}%
_{\tilde{\nu}}$, respectively, except that their indices are restricted to $%
1,\;2,\;$and $3.$ The smallest representation, with $\widetilde{n}_{1}+%
\widetilde{n}_{2}+\widetilde{n}_{3}\equiv \widetilde{N}^{\prime }=0$, is the 
\textbf{1} consisting of $\;\tilde{\phi}_{000}=\exp \{-\frac{1}{2}\tilde{%
\beta}^{2}[(\tilde{x}^{1})^{2}+(\tilde{x}^{2})^{2}+(\tilde{x}%
^{3})^{2}]\}\equiv \tilde{F}^{\prime }$. The next smallest, with $\widetilde{%
N}^{\prime }=1,$ is the \textbf{3} consisting of\ $\tilde{\phi}_{100}=\tilde{%
x}^{1}\tilde{F}^{\prime },\;\tilde{\phi}_{010}=\tilde{x}^{2}\tilde{F}%
^{\prime },$ and $\tilde{\phi}_{001}=\tilde{x}^{3}\tilde{F}^{\prime }.$

\end{document}